\documentclass[fleqn,10pt]{wlscirep}
\title{Dynamic scaling for the growth of non-equilibrium fluctuations during thermophoretic diffusion in microgravity}

\author[1]{Roberto Cerbino}
\author[2]{Yifei Sun}
\author[2]{Aleksandar Donev}
\author[3,*]{Alberto Vailati}
\affil[1]{Universit\`a degli Studi di Milano, Dipartimento di Biotecnologie Mediche e Medicina Traslazionale, Milan, I-20133, Italy}
\affil[2]{New York University, Courant Institute of Mathematical Sciences, New York, NY 10012, USA}
\affil[3]{Universit\`a degli Studi di Milano, Dipartimento di Fisica, Milan, I-20133, Italy}
\affil[*]{alberto.vailati@unimi.it}

\begin{abstract}
Diffusion processes are widespread in biological and chemical systems, where they play a fundamental role in the exchange of substances at the cellular level and in determining the rate of chemical reactions. Recently, the classical  picture that portrays diffusion as random uncorrelated motion of molecules has been revised, when it was shown that giant non-equilibrium fluctuations develop during diffusion processes.  Under microgravity conditions and at steady-state, non-equilibrium fluctuations exhibit scale invariance and their size is only limited by the boundaries of the system. In this work, we investigate the onset of non-equilibrium concentration fluctuations induced by thermophoretic diffusion in microgravity, a regime not accessible to analytical calculations but of great relevance for the understanding of several natural and technological processes.  A combination of state of the art simulations and experiments allows us to attain a fully quantitative description of the development of fluctuations during  transient diffusion in microgravity. Both experiments and simulations show that during the onset the fluctuations exhibit scale invariance at large wave vectors. In a broader range of wave vectors simulations predict a spinodal-like growth of fluctuations, where the amplitude and length-scale of the dominant mode are determined by the thickness of the diffuse layer.
\end{abstract}
\begin{document}

\flushbottom
\maketitle

\thispagestyle{empty}

\section*{Introduction}

Diffusion in liquid mixtures and suspensions represents a fundamental spontaneous mass transfer mechanism at the microscopic scale. For instance, it regulates transport processes in the cell, the growth of crystals and the kinetics of aggregation of macromolecules and colloidal particles in suspension. During the last 20 years it has been shown, both theoretically and experimentally \cite{booksengers06}, that diffusion is accompanied by non-equilibrium concentration fluctuations exhibiting generic scale invariance \cite{grinstein95} in the length scale range from the molecular scale up to the macroscopic size of the system. Quite interestingly, it has been shown that non-equilibrium fluctuations do not represent merely a perturbation of a macroscopic state; instead, the diffusive flux can be understood to be entirely generated by non-equilibrium fluctuations \cite{brogioli00,donev11,donev14}. On Earth the scale invariance of the fluctuations is broken at small wave vectors by the presence of the gravity force \cite{segre93} that either quenches \cite{vailati96,vailati97} or amplifies \cite{wu95, ahlers, giavazzi09} long wavelength fluctuations, depending on whether the density profile is stabilizing or not. Under microgravity conditions the scale invariance is broken by the finite size of the diffusing system \cite{vailati2011}.

So far, theoretical models suitable to describe the statistical properties of non-equilibrium fluctuations have been developed only for systems at steady state \cite{booksengers06} or for systems whose macroscopic state evolves much slower than the fluctuations \cite{vailati98}. While this is always the case for weakly confined systems undergoing diffusion on Earth, the situation in microgravity conditions is more complex and no theoretical model is currently able to provide an analytic description of non-equilibrium fluctuations occurring during transient diffusion processes in microgravity. This is due to the fact that under microgravity conditions the modes associated to the macroscopic state and to the fluctuations evolve with the same timescales, thus preventing the separation of the two contributions. However, since the advent of space platforms, several experiments controlled by diffusion have been performed in a microgravity environment, which guarantees the absence of spurious convective motions. Notable examples include experiments on the crystallization of proteins \cite{proteins,snell05}, critical phenomena \cite{barmatz07, beysens14}, the investigation of the influence of vibration on diffusion \cite{shevtsova10, shevtsova11}, and of transport properties in ternary mixtures \cite{shevtsova14}. Therefore, the understanding of the onset of concentration fluctuations during diffusion in the absence of gravity represents an important feat both from the fundamental point of view, due to the lack of suitable theoretical models, and from the experimental point of view, due to the huge investment required to perform experiments in Space.

\begin{figure} [ht]
\centering\includegraphics [width=0.5\linewidth]{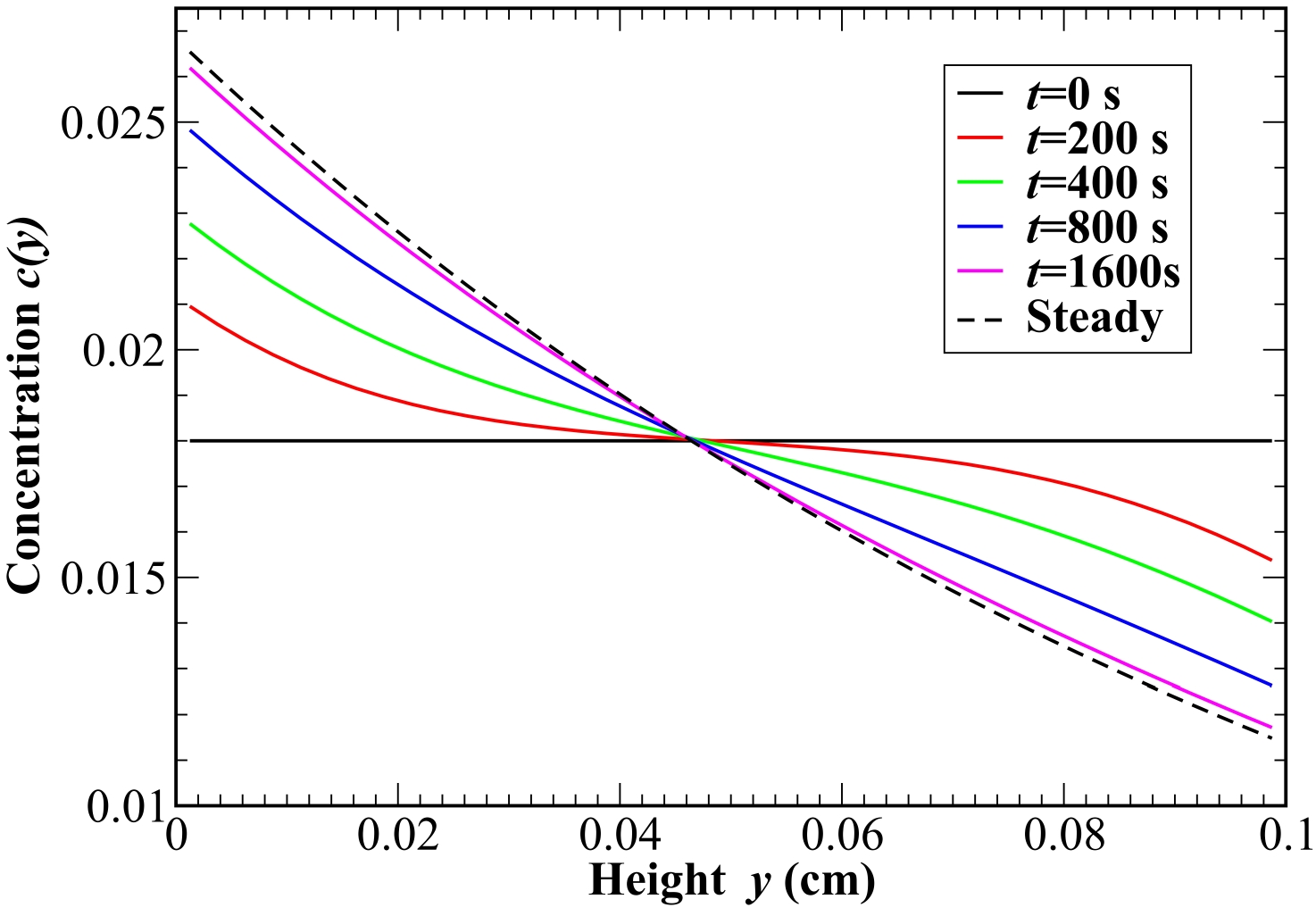}
\caption{Numerically calculated time evolution of the concentration profile}
\label{fig concprof}
\end{figure}

In this work, we investigate both experimentally and computationally the onset of non-equilibrium concentration fluctuations in a polymer suspension under microgravity conditions. We quickly apply a temperature gradient to the initially homogeneous polymer solution. The gradient gradually induces the formation of a concentration profile through thermophoresis (Fig. \ref{fig concprof}) \cite{bookdegroot62}. The time evolution of the fluctuations is monitored experimentally by using a quantitative shadowgraph technique \cite{booksettles01,trainoff02}. The fluctuations are also simulated under the same conditions by using a finite-volume method recently developed for the study of giant fluctuations in confinement \cite{MultiscaleIntegrators,LLNS_Staggered}. For large wave vectors, the scale invariance of the fluctuations is confirmed,  both by experiments and simulations, also during the transient. Interestingly, simulations predict that a dominant mode in the structure factor of the fluctuations is found at small wave vectors during transient diffusion. The wave vector $k_{m}$ associated to this dominant mode decreases as time goes by, with a kinetics compatible with a diffusive growth. For long time, the peak disappears and is replaced by the expected plateau due to the effect of the impermeable boundaries \cite{booksengers06,dezarate04}. The kinetics observed during the transient bears many similarities with that of spinodal decomposition \cite{spinodalhuang}, the most notable feature being that the structure factor $S(k,t)$ of the fluctuations at different times $t$ can be scaled onto a single master curve $F(k/k_{m})$ by using a scaling relation $S(k/k_{m},t)=k_{m}(t)^{-\alpha} F(k/k_{m})$ \cite{spinodalbinder,spinodalmarro,spinodalfurukawa}.

\section*{Results}

\begin{figure} [ht]
\centering\includegraphics[width=\linewidth]{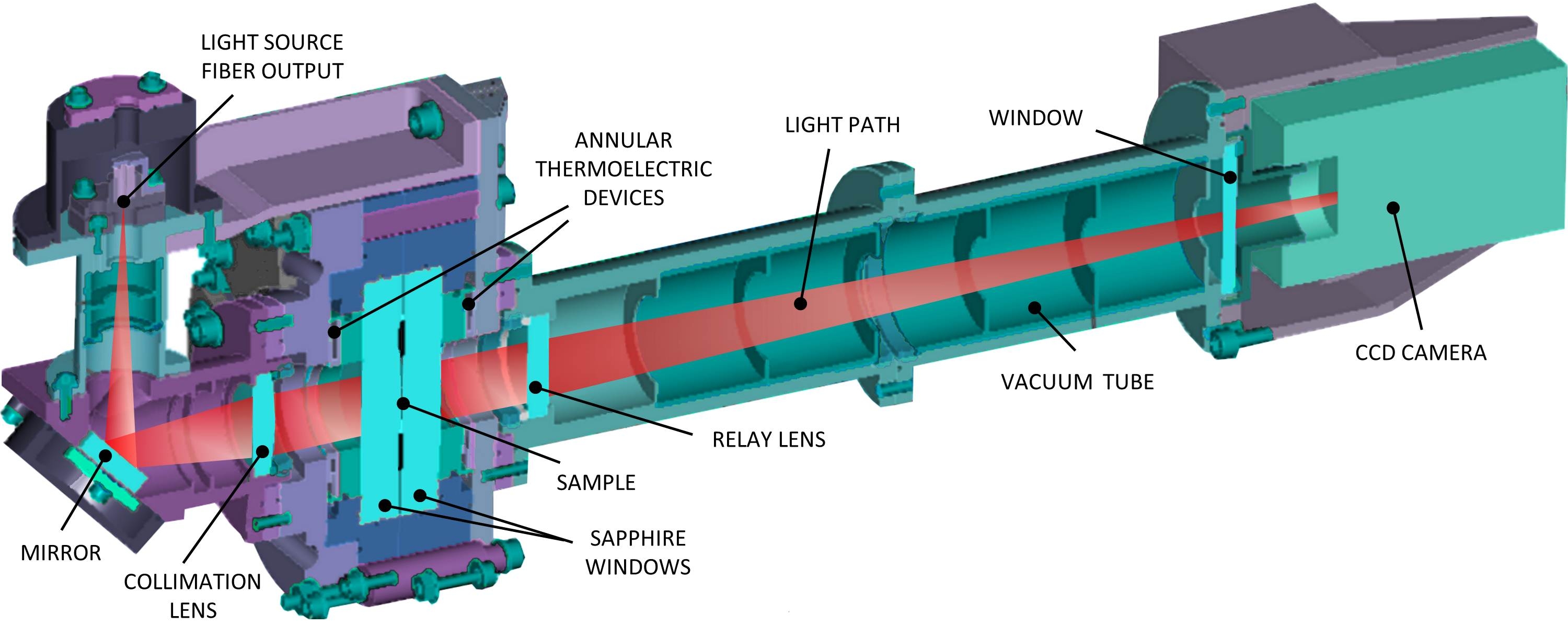}
\caption{GRADFLEX setup. The sample is sandwiched between two sapphire windows in thermal contact with thermo-electric devices. The light coming from a Light Emitting Diode through a fiber is steered by a mirror and collimated onto the sample by a lens. A relay lens collects the light from the main beam and the light scattered by the sample, which interfere onto the sensor of a CCD camera. The light path is kept under vacuum.  }
\label{fig setup}
\end{figure}

Experiments have been performed aboard the FOTON M3 spaceship by using the GRADFLEX facility developed by ESA \cite{vailati2011,takacs2011}. Foton M3 is an unmanned spaceship orbiting at an average distance from the Earth of the order of 300 km. The great advantage of such a platform with respect to other facilities, such as the International Space Station, is the very low level of residual gravity, of the order of 0.7 $\mu$g on average. The GRADFLEX setup comprises a thermal gradient cell and a quantitative shadowgraph optical diagnostics (Fig. \ref{fig setup}). The sample is a suspension of polystyrene (molecular weight 9100) in toluene with a weight fraction concentration of 1.8\%. It is contained inside the thermal gradient cell \cite{vailati06,vailati2011} whose thermal plates are two sapphire windows. These windows are in thermal contact with two annular thermo-electric devices. This peculiar configuration of the cell enables using the sapphire windows both as thermal plates and as observation windows for the detection of fluctuations. The temperature of the sapphire windows is monitored by using thermistors that drive two Proportional-Integral-Derivative servo control loops. The light source is a superluminous Light Emitting Diode coupled to an optical fiber. The collimated light coming from the diode crosses the sample, where it gets partially scattered by non-equilibrium fluctuations. The superposition of the scattered light and of the main beam gives rise to an interference pattern onto the sensor of a Charged Coupled Device camera. This pattern can be analyzed statistically by using the theory of quantitative shadowgraphy \cite{booksettles01,trainoff02} to determine the structure factors of temperature and concentration non-equilibrium fluctuations. A typical measurement run performed in space involves the automated execution of a stabilization phase followed by a measurement phase: after a stabilization of the equipment lasting 3 hours, the sample is kept at a uniform temperature $T=30^o \mathrm{C}$ for 90 minutes; the quick imposition of a temperature difference $\Delta T$ at time $t=0$ determines the start of the diffusive process. The typical time constant associated to the growth of temperature difference across the sample is of the order of $\tau_{T} \approx 100s$, significantly smaller than the time needed for the diffusive process to reach a steady state $\tau_{c}\approx 2000s$. The presence of a temperature gradient inside the sample determines a non-equilibrium thermophoretic contribution to the average mass flux $j=-\rho D[\nabla c+ S_{T}c(1-c)\nabla T]$ that gives rise to the development of an almost exponential concentration profile (Fig. \ref{fig concprof}) \cite{ruckenstein81}. Here $D=1.97 \times 10^{-6}\mathrm{cm^2/s}$ is the mass diffusion coefficient and $S_{T}=6.49 \times 10^{-2} \mathrm{K}^{-1}$ is the Soret coefficient. At steady state and at the impermeable boundaries the net flux must vanish and the concentration profile is characterized by a gradient $\nabla c=-S_{T}c(1-c)\nabla T$.

\subsection*{Non-equilibrium temperature and concentration fluctuations at steady state}

\begin{figure} [ht]
\centering\includegraphics[width=0.75\linewidth]{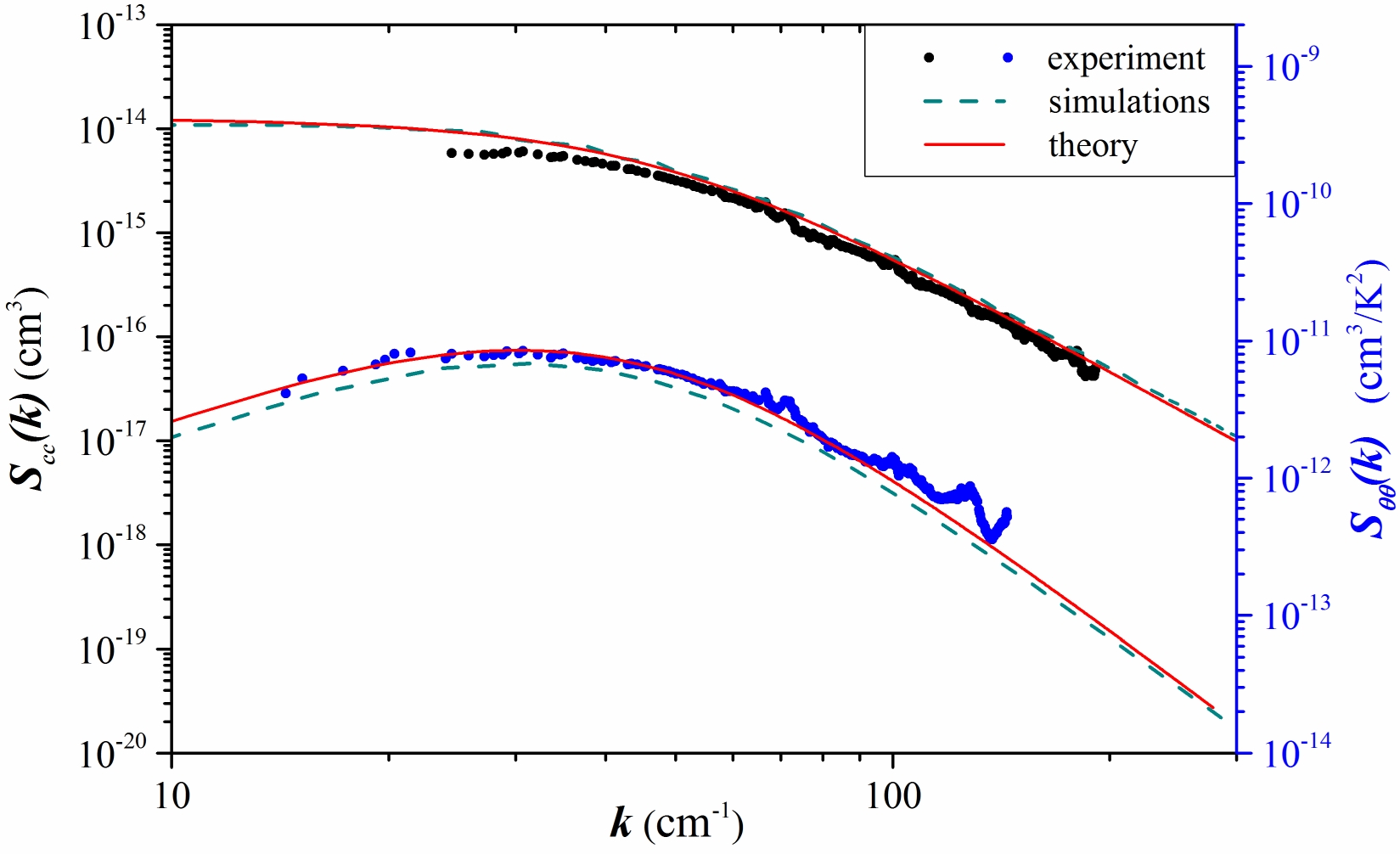}
\caption{Structure factors of the non-equilibrium temperature (data on bottom and right y-axis) and concentration (data on top and left y-axis) fluctuations at steady state under microgravity conditions. Circles: experimental results; dashed lines: simulations; solid lines: theory \cite{dezarate06, ortizpersonal}.}
\label{fig steady state}
\end{figure}

The simultaneous presence of a temperature and a concentration gradient determines the onset of both temperature and concentration fluctuations. For the polymer suspension of interest here the diffusion coefficient is much smaller that the thermal diffusivity $\kappa=8.95 \times 10^{-4}\mathrm{cm^2/s}$ of the sample, and the timescales for the relaxation of temperature and concentration fluctuations are well separated. The wide difference between these timescales was used in Ref. \citenum{vailati2011} to estimate the Fourier power spectrum of concentration fluctuations by using a standard dynamic analysis \cite{croccolo07,vailati2011}. Here we use a refined procedure that allows obtaining also the power spectrum of the temperature fluctuations. In addition, we use the power spectra of both temperature and concentration fluctuations to estimate the corresponding structure factors at steady state (Fig. \ref{fig steady state}). The advantage of this procedure lies in the fact that it allows a precise determination of the temperature difference $\Delta T$ across the sample, which was not measured directly in the GRADFLEX experiment but rather estimated from thermal modeling of the sample cell. Fitting (bottom dashed line in Fig. \ref{fig steady state}) the temperature $S_{\theta\theta}(k)$ to the analytical theoretical expression determined by De Zarate and Sengers by using a Galerkin approximation \cite{dezarate04,booksengers06} provides the estimate $\Delta T=13.25 \mathrm{K}$, which is 24$\%$ smaller than what was previously estimated  by thermal modeling \cite{vailati2011}. The Galerkin approximation systematically under-estimates the structure factor at small wave numbers \cite{dezarate06} by  a factor of $500.5/720=0.695$ and is therefore a source of additional error; the computational method used here does not make any such uncontrolled approximations. There is presently no exact closed-form theoretical expressions available for perfectly conducting boundaries. 

Once a reliable estimate for $\Delta T$ has been obtained, we applied a similar procedure to obtain the structure factor of concentration fluctuations at steady state in absolute units. The experimental estimate turns out to be systematically slightly smaller than the theoretical predictions made by using a recent exact prediction \cite{ortizpersonal} (solid line in Fig. \ref{fig steady state}). We believe that this discrepancy can be attributed to an actual concentration of the sample about 10 \% below the nominal value of $c=1.8$ $\%$ w/w. The results of simulations are also shown in Fig. \ref{fig steady state} (dashed lines) and at large wave vectors are in fair agreement with both experiments and theory. In Fig. \ref{fig steady state}, it can be noticed that we could not obtain experimental results at very small wave vectors. This is due to the presence of a drift of the optical background of the shadowgraph setup for long times, which prevents the characterization of the concentration fluctuations at small wave vectors, but in principle does not affect much the short-lived temperature fluctuations.

At large wave vectors, the structure factors of both temperature and concentration fluctuations scale as $k^{-4}$, mirroring the scale invariance of the fluctuations. However, at a wave vector  $k_{\text{fs}}\approx\pi/h$, the finite thickness $h$ of the sample along the applied gradient produces different effects on the two structure factors because of the different boundary conditions for concentration and temperature. Indeed, the boundaries are impermeable to mass but conduct heat very well. As a consequence, long wave length temperature fluctuations can be dissipated effectively through the boundaries and a peak in the temperature $S_{\theta\theta}(k)$ can be observed. In contrast, in the case of concentration fluctuations the boundaries are impermeable and long wavelength fluctuations can be dissipated by diffusion only, which leads to a plateau in $S_{cc}(k)$ for $k<k_{\text{fs}}$ \cite{dezarate04,booksengers06}.

\subsection*{Onset of non-equilibrium concentration fluctuations}
The selected experimental sample represents an ideal system to investigate the onset of concentration fluctuations. In fact, the small diffusion coefficient determines the progressive development of a macroscopic concentration profile lasting about 30 minutes. The sample is initially kept at a uniform temperature of $30^o$C.  The diffusion process is started by imposing a temperature difference $\Delta T=13.25$ K at $t=0$. Every $10$ s we record a shadowgraph image of the sample. The long timescale associated to the development of a macroscopic concentration profile enables us to grab $200$ shadowgraph images of the sample during the approach to steady state. Due to the fact that the system is evolving in time during the transient, it is not possible to recover the structure factors of non-equilibrium fluctuations during the transient by applying the same procedure  used to recover them at steady state. Instead, in this case we rely on a dedicated processing algorithm that takes advantage of the fact that after about $100$ s the temperature profile reaches a steady state. For this reason, the first $190$ s of the process have not been included in the analysis. Starting from the image taken at $t=200$ s, structure factors of the concentration fluctuations have been averaged on groups of $10$, $15$, $20$, $30$, $80$ images, corresponding to average times of $245$, $370$, $545$, $1345$ s. This procedure allows reducing the noise on the structure factor by increasing the statistical sample, without losing much temporal resolution.

\begin{figure} [ht]
\centering\includegraphics [width=0.75\linewidth]{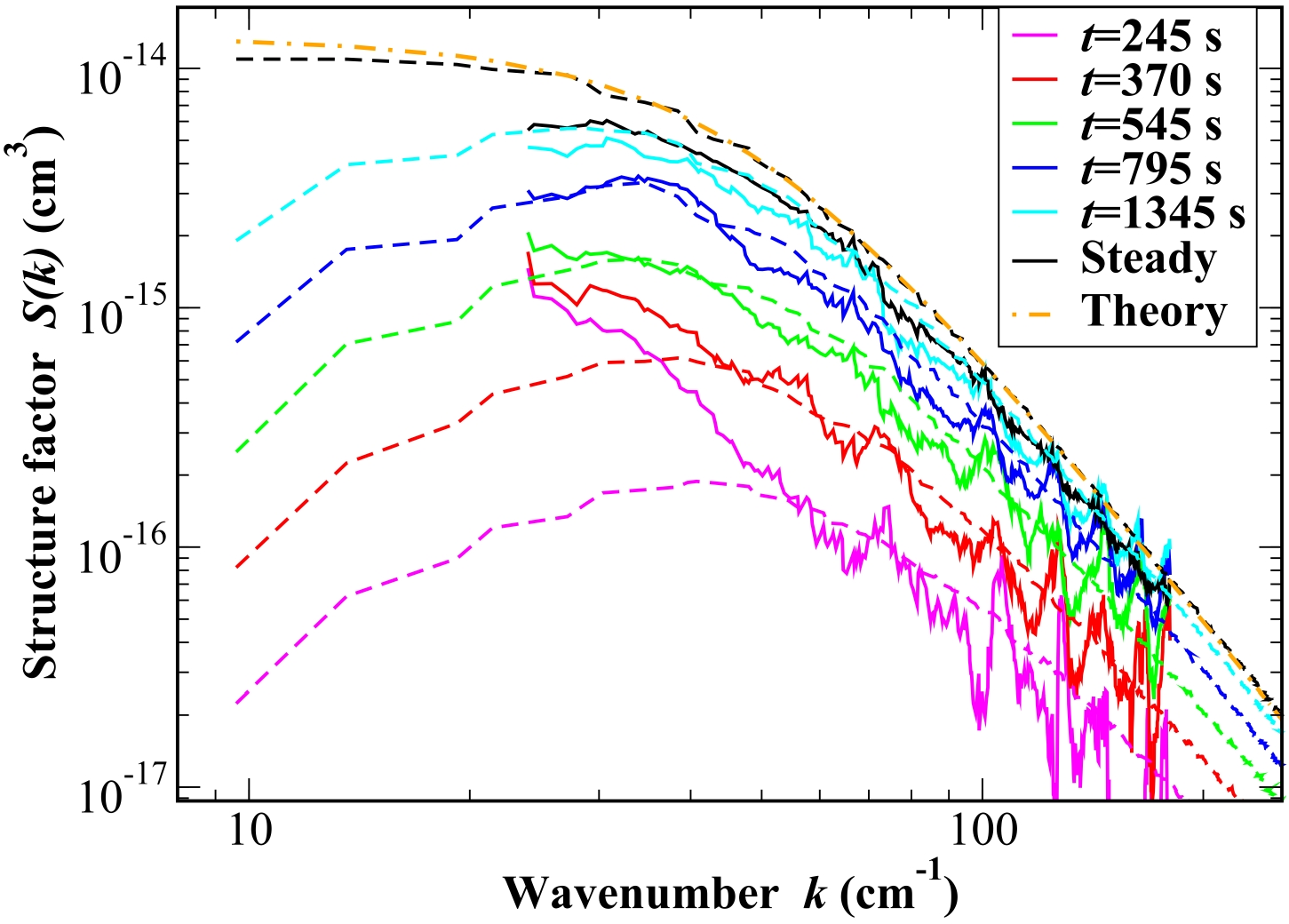}
\caption{Time evolution of the structure factor of non-equilibrium concentration fluctuations during the approach to steady state under microgravity conditions. Solid lines: experimental data; dashed lines: simulations; dashed-dotted line: exact theory \cite{ortizpersonal}}
\label{fig transient}
\end{figure}

\begin{figure} [ht]
\centering\includegraphics[width=\linewidth]{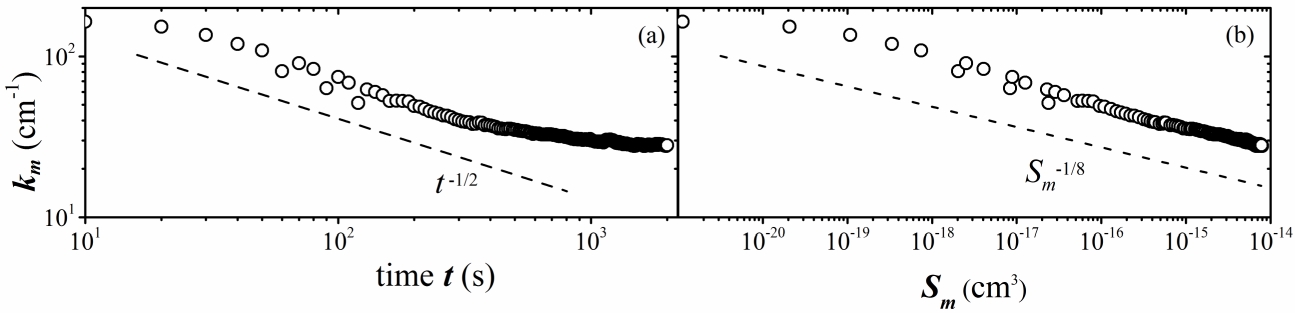}
\caption{a) Time evolution of the wave vector of the dominant mode. The dashed line corresponds to a diffusive behavior $k_{m} \propto (D t)^{-1/2}$. b) Wave vector of the dominant mode as a function of the mean squared amplitude (power) of the mode. The dashed line correspond to a power law behavior $k_{m} \propto S_{m}^{-1/8}$}
\label{fig kinetics}
\end{figure}

Due to the lack of a theoretical model suitable to deal with a transient system, we have performed simulations under conditions and sampling procedure mirroring those found in the experiment. A comparison of the experimental and simulated data is shown in Fig. \ref{fig transient}. The experimental results are in fair agreement with those of the simulations, the only exception being the small $k$ behavior of the structure factors corresponding to $245$ s and $370$ s. For such times, an effective subtraction of the optical background is difficult due to the sudden application of the temperature difference, which is particularly limiting when the signal is weakest. To partially avoid these disturbances the optical path is kept under vacuum, but when the light scattered by the fluctuations is weak the signal at small wave vector is dominated by fluctuations in the optical path of the probe beam and by mechanical drifts of the setup. This effect limits our accessible range and prevents the experimental observation of a peak in the structure factors, which is well visible in the simulation results only during the short-time transient. This peak is associated to the presence of a dominant mode with a wave vector that gradually decreases in time (Fig. \ref{fig kinetics}a), while  the amplitude of the mode increases progressively (Fig. \ref{fig kinetics}b). 

A first understanding of the presence of a peak can be achieved by taking into consideration that in the presence of fully developed temperature and concentration profiles the structure factor $S(k,t)$ of concentration fluctuations grows diffusively. In fact, under these conditions it can be shown that, ignoring the influence of the boundaries, $S(k,t) \propto [1-\exp(-2 D k^2 t)] S(k,\infty)$ where $S(k,\infty)$ is the structure factor at steady state. A simple model along this lines provides the right qualitative behavior and gives rise to a peak in the structure factor behaving asymptotically as $k^2$ and $k^{-4}$ at small and large $k$, respectively. However, in any real diffusive process the modes associated to fluctuations and to the macroscopic state evolve with the same time constants. Therefore, the assumption of an initial fully developed concentration gradient on top of which fluctuations develop is rather unrealistic. In practice, real effects like the progressive development of a temperature gradient and the subsequent growth of boundary layers in the concentration profile are difficult to model theoretically, but can be modeled exactly by means of simulations. The fit of the peak of the structure factors of the simulated fluctuations allows us to recover the wave vector $k_m$ and the structure factor $S_m=S(k_m)$ of the dominant mode. The time evolution of $k_m$ at times smaller than about $100$ s is compatible with a diffusive growth of the mode $k_{m} \propto (D t)^{-1/2}$ (Fig. \ref{fig kinetics}a). During this initial phase, the two boundary layers grow without feeling much the presence of each other. However, after a time $\tau= (h/2)^2/(\pi^2 D)\approx 120s$ they reach a thickness comparable to $h/2$, and the system enters a diffusive regime where finite size effects become relevant, as mirrored by the slowing down of the decrease of $k_m$. 

\subsection*{Dynamic scaling of non-equilibrium concentration fluctuations}

An important feature of the dynamics of the dominant mode is the relation between $k_m$ and  $S_m$ (Fig. \ref{fig kinetics}b). One can appreciate that at the times larger than 200 s, when the system has entered the restricted diffusion regime, $k_m$ and $S_m$ are related by a power law $k_{m} \propto S_{m}^{\beta}$, with exponent $\beta=0.12$. 

\begin{figure} [ht]
\centering\includegraphics[width=0.75\linewidth]{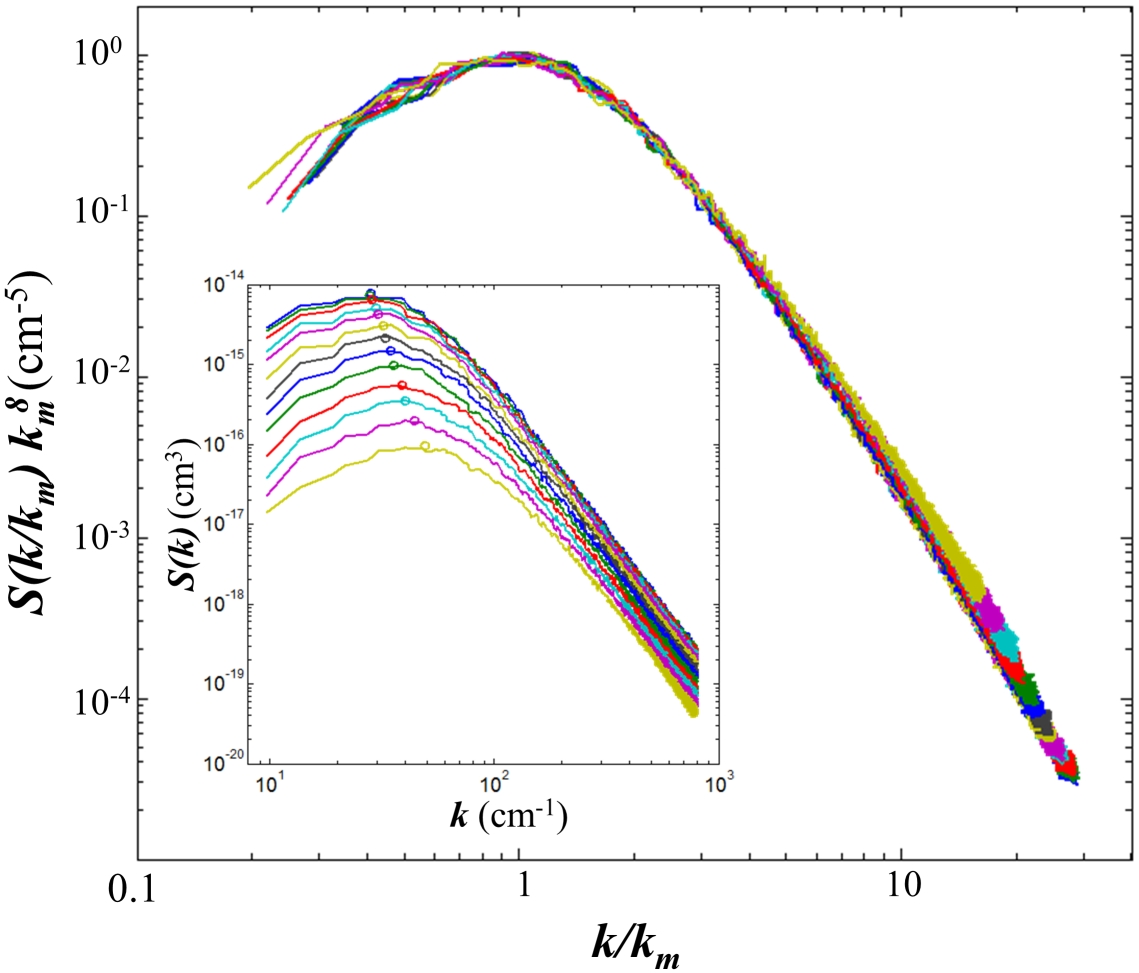}
\caption{ Dynamic scaling of the spectra of the non-equilibrium fluctuations during the approach to steady state. The inset shows the unscaled structure factors. Times span the range $200s\leqq t \leqq2000s$ and are distributed geometrically with a multiplier of 1.21 (for a total of 13 curves).}
\label{fig scaling}
\end{figure} 

Qualitatively, this behavior is similar to that reported for spinodal decomposition \cite{spinodalhuang,spinodalmarro,spinodalbinder,spinodalfurukawa} and other phenomena, such as colloidal aggregation \cite{spinodalcarpineti}. The growth dynamics of the structure factor of the concentration perturbations associated to these phenomena is such that the structure factor exhibits dynamic scaling $S(k/k_{m}, t)=k_{m}^{-\alpha} F(k/k_{m})$, where $F(k,k_{m})$ is a time independent master curve and $\alpha$ a power law exponent, which in the case of spinodal decomposition corresponds to the dimensionality of the space. This suggests that our results are compatible with a scaling law akin to that of spinodal decomposition with a power law exponent $\alpha=1/\beta \approx 8$. By scaling the structure factors of simulations in the time range $200s \leq t \leq 2000s$ using the relation$S(k/k_{m}, t) k^{8} =F(k/k_{m})$ we get that the curves nicely collapse onto a single, time-independent, master curve $F(k/k_{m})$ (Fig. \ref{fig scaling}).

\section*{Discussion}

It turns out that there are some qualitative analogies between the growth of non-equilibrium concentration fluctuations and that of the domains in spinodal decomposition. In the case of spinodal decomposition the presence of a dominant mode is due to the fact that the process is controlled by a generalized diffusion equation where the diffusion coefficient is negative. This uphill diffusion determines the growth of the domains and the progressive buildup of large concentration gradients. In the case of non-equilibrium fluctuations we know that in microgravity both the macroscopic state and the fluctuations are controlled by a diffusion equation with positive $D$ in the presence of a steady counter-flux determined by the Soret Effect. At steady state the diffusive and Soret fluxes balance each other and there's no net mass flow through the sample. However, during the transient the mass flux is dominated by the Soret contribution, and the net balance in the flux of mass determines the growth of a concentration gradient, similarly to what happens during the demixing process that drives spinodal decomposition. For spinodal decomposition the power law exponent used for the scaling is the dimensionality of the space; in our case it is close to 8.

Our results provide experimental evidence that linearized fluctuating hydrodynamics quantitatively describes the time-dependent growth of fluctuations during {\em transient} diffusion processes. Our experimental results are calibrated and compared to theory in absolute units, thus significantly extending previous studies for {\em steady-state} fluctuations in microgravity. Analytical calculations are essentially infeasible in the presence of a transient reference state, especially in the absence of separation of time scales as in diffusive mixing in microgravity. The development of numerical techniques for solving the equations of fluctuating hydrodynamics \cite{MultiscaleIntegrators,LLNS_Staggered} has allowed us to predict the existence of a dynamic scaling law during the development of non-equilibrium fluctuations that has not yet been observed in experiments.

\section*{Methods}

\subsection*{Measurement setup}
The GRADFLEX Mixture setup comprises a thermal gradient cell and a shadowgraph optical diagnostics. The gradient cell \cite{vailati06,vailati2011} consists of two - 12 mm thick - sapphire windows kept at a distance of 1.00 mm from each other by means of a calibrated spacer. The lateral confinement of the sample is achieved by means of a Viton gasket with an inner diameter of 27 mm. The relatively high thermal conductivity of sapphire guarantees a temperature uniform within 3\% across the contact surface of the window with the sample. Each sapphire window is sandwiched with a coupling ring made of aluminum that brings it into thermal contact with an annular thermo-electric device with an inner bore with a diameter of 27 mm. The temperature of the sapphire windows is monitored by Negative Temperature Coefficient thermistors that drive two independent Proportional-Integral-Derivative servo-controls, which allow to achieve a stability of the temperature of the windows the order 10 mK over 24 hours. The optical shadowgraphy diagnostics makes use of a super-luminous Light Emitting Diode with a wavelength of 680 nm and a bandwidth of 13 nm. The LED is coupled to a mono-mode optical fiber. The diverging beam coming out of a fiber is steered by a mirror and collimated by an achromatic doublet. The role of the steering mirror is to fold the optical path, to maintain the size of the instrument compact. The collimated beam goes through the sample, where it gets partially scattered by non-equilibrium fluctuations. The main beam and the scattered light are collected by a relay lens and superimposed onto the sensor of a Charged Coupled Device camera with a resolution of $1024\times1024$ pixel and a pixel depth of 10 bit, which records an image every 10s . In order to avoid disturbances generated by air, the light path is kept under vacuum by means of vacuum tube which can be connected to the outer environment of the spaceship by means of a remotely actuated valve.

\subsection*{Optical diagnostics}

\subsubsection*{Quantitative shadowgraphy}
The non-equilibrium temperature and concentration fluctuations arising as a consequence of the application of a macroscopic temperature gradient to a polymer solution (from here-on the sample) give rise to refractive index fluctuations that can be detected by using optical shadowgraphy \cite{booksettles01,trainoff02,vailati06,vailati2011}. The phase of a plane wave of intensity $I_{0}$ that impinges on the sample is locally altered by any refractive index inhomogeneity, which causes light scattering. Quantitative shadowgraphy is based on the idea that, sufficiently far away from the sample and for weakly scattering systems, the scattered light interferes with the transmitted plane wave creating a time-dependent hologram $I(x,y,t)\simeq I_{0}+2\sqrt{I_{0}}\operatorname{Re}[E_{s}(x,y,t)]$ at some distance $z$ from the sample\cite{cocis}. The shadowgraph signal is defined as $s(x,y,t)=[I(x,y,t)/I_{0}]-1=2\operatorname{Re}[E_{s}(x,y,t)/E_{0}]$, where $E_s(x,y,t)$ is the amplitude of the electric field scattered from the sample at distance $z$. If we indicate with $s(\vec{k},t)$ the spatial two-dimensional Fourier transform of $s(\vec{x},t)$ (from hereon $\vec{k}=(k_{x},k_{y})$ and $\vec{x}=(x,y))$, then the Fourier power spectrum of the shadowgraph signal is given by \begin{equation} \label{eq:pspec}
A(\vec{k})=\left\langle\left|s(\vec{k},t)\right|^{2}\right\rangle_{t}=4VT_{F}(\vec{k})k_{0}^{2}\left[\left(\frac{\partial n}{\partial c}\right)^{2}S_{cc}(\vec{k})+\left(\frac{\partial n}{\partial T}\right)^{2}S_{\theta\theta}(\vec{k})\right]\doteq A_{cc}(\vec{k})+A_{\theta\theta}(\vec{k})
\end{equation} where $V$ is the imaged volume, $k_{0}$ is the wave-vector of the incident light, $n(c,T)$ is the refractive index, $T_{F}$ is the transfer function of shadowgraphy, and $S_{cc}$ and $S_{\theta\theta}$ are the structure factors of concentration and temperature fluctuations as defined in Ref. \citenum{booksengers06}, respectively. Knowledge of the transfer function $T_{F}(\vec{k})$ is thus needed for the quantitative determination of the structure factors of the fluctuations and it requires a suitable calibration of the optical setup \cite{cerbino08, vailati2011}.

\subsubsection*{Transfer function calibration}
The optical setup was calibrated by using polystyrene spheres with a nominal diameter of $2.0$ $\mu$m, dispersed in isopropyl alcohol \cite{vailati2011}. Such sample provides a large optical contrast that in the wave-vector range accessible to our experiments gives rise to a constant scattering intensity, representing thereby the ideal calibration sample. The amplitude $A_{cal}(\vec{k})$ determined by using the calibration sample was fitted to the function $A_{cal}(\vec{k})=U(\vec{k})\sin^2{\left[k^2z/(2k_{0})+\phi\right]}+V(\vec{k})$, where $k=\sqrt{k^2_{x}+k^2_{y}}$ and where $\phi=1.78$ was found to match expectations from Mie theory \cite{vailati2011}. The so determined $U$ and $V$ were thus used to reconstruct the transfer function $T_{F}$ for the fluctuations that is obtained when $\phi=0$ is used. Once the transfer function $T_{F}(\vec{k})$ is known, the setup can be used for the quantitative assessment of the static and dynamic scattering properties of the sample.

\subsection*{Reduction of experimental results}

\subsubsection*{Dynamic analysis and isolation of the concentration and temperature contributions at steady state}
A typical analysis of the shadowgraph images $I(x,y,t)$ acquired at steady state at various times $t$ involves the processing of $5000$ images. By using a variant of the differential dynamic algorithm the image structure function $D_{I}(\vec{k},\Delta t)=\left\langle \left|I(\vec{k},t+\Delta t)-I(\vec{k},t)\right|^{2}\right\rangle_{t}$ is calculated by averaging over pairs of images separated by the same $\Delta t$ \cite{croccolo07,java}. Here $I(\vec{k},t)$ is the spatial two-dimensional Fourier transform of $I(\vec{x},t)$. Theoretical expectation is that $D_{I}(\vec{k},\Delta t)=2A(\vec{k})\left[1-f(\vec{k},\Delta t)\right]+2B(\vec{k})$, where $A(\vec{k})$ is given in Eq. \ref{eq:pspec}, $B(\vec{k})$ is a dynamic background term that accounts for the noise in the detection chain and $f(\vec{k},\Delta t)$ is the intermediate scattering function of the fluctuations. For our experiments $f(\vec{k},\Delta t)=\frac{A_{cc}(\vec{k})}{A(\vec{k})}e^{-\frac{t}{\tau_{c}(\vec{k})}}+\frac{A_{\theta\theta}(\vec{k})}{A(\vec{k})}e^{-\frac{t}{\tau_{T}(\vec{k})}}$, where $\tau_{c}(\vec{k})$ and $\tau_{T}(\vec{k})$ are the characteristic correlation times of the concentration and temperature fluctuations, respectively. One thus has $D_{I}(\vec{k},\Delta t)=2\left[A_{cc}(\vec{k})\left(1-e^{-\frac{t}{\tau_{c}(\vec{k})}}\right)+A_{\theta\theta}(\vec{k})\left(1-e^{-\frac{t}{\tau_{T}(\vec{k})}}\right)\right]+2B(\vec{k})$. In practice, in our k-range $\tau_{T}(\vec{k})$ is smaller than the time elapsed between the acquisition of two successive images (10 s). As a result, temperature fluctuations appear as uncorrelated background signal and their static scattering contributes to the background. One has $D_{I}(\vec{k},\Delta t)=2A_{cc}(\vec{k})\left(1-e^{-\frac{t}{\tau_{c}(\vec{k})}}\right)+2B_{eff}(\vec{k})$, where the effective background $B_{eff}(\vec{k})=B(\vec{k})+A_{\theta\theta}(\vec{k})$ incorporates also the static scattering signal $A_{\theta\theta}(\vec{k})$ from temperature fluctuations. Fitting of the experimental curves for $D_{I}(\vec{k},\Delta t)$ provides thus estimates for $A_{cc}(\vec{k})$, $B_{eff}(\vec{k})$ and $\tau_{c}(\vec{k})$. We also independently determined $B(\vec{k})$ from the dynamic analysis of images acquired in the absence of any temperature and concentration gradients, which in turn enabled us obtaining also an estimate for $A_{\theta\theta}(\vec{k})$. Using Eq. \ref{eq:pspec}, we recovered the structure factors $S_{cc}(\vec{k})$ and $S_{\theta\theta}(\vec{k})$ at steady state (Fig. \ref{fig steady state}).

\subsubsection*{Analysis of the transient}
The 200 shadowgraph images acquired during the transient contain contributions coming from concentration fluctuations, temperature fluctuations, dynamic noise and static background. The Fourier power spectrum $B(\vec{k})$ of the dynamic background is time independent and can be characterized accurately from the dynamic analysis at steady state, as described above. Similarly, after a time of about 100s needed for the onset of temperature fluctuations, their contribution $A_{\theta\theta}(\vec{k})$ to the Fourier power spectrum of the shadowgraph signal becomes time independent and coincides with that  determined at steady state.  Conversely, the static background contribution represents a time-independent additive term to each shadow image arising from a non-uniform illumination of the sample. This contribution could be eliminated easily by using the differential dynamic analysis described above. However, the differential dynamic analysis cannot be applied to the images taken during the transient due to the  limited statistical sample. To overcome this limitation, we determined the static background term by averaging in time 700 images collected at steady state $I_{SB}(x,y)=\langle I(x,y,t)\rangle$. The shadowgraph signal during the transient is then defined as $s_{tr}(x,y,t)=[I(x,y,t)/I_{SB}(x,y)]-1$ and its Fourier power spectrum is given by $A_{tr}(\vec{k})=\left\langle\left|s_{tr}(\vec{k},t)\right|^{2}\right\rangle_{t}=A_{cc,tr}(\vec{k})+A_{\theta\theta}(\vec{k})+B(\vec{k})$.
The temporal average was processed by skipping the first 19 images, to avoid effects related to the onset of temperature fluctuations, and by averaging the following images in groups of $10$, $15$, $20$, $30$, $80$, corresponding to average times of $245$, $370$, $545$, $1345$ s, respectively. The structure factor of transient concentration fluctuations in absolute units was then be determined from the relation $S(\vec{k})=\left[A_{tr}(\vec{k})-A_{\theta\theta}(\vec{k})-B(\vec{k})\right]/\left[4VT_{F}(\vec{k})k_{0}^{2}\left(\frac{\partial n}{\partial c}\right)^{2}\right]$.

\subsection*{Numerical simulations}

\global\long\def\V#1{\boldsymbol{#1}}
\global\long\def\M#1{\boldsymbol{#1}}
\global\long\def\Set#1{\mathbb{#1}}

\global\long\def\D#1{\Delta#1}
\global\long\def\d#1{\delta#1}

\global\long\def\norm#1{\left\Vert #1\right\Vert }
\global\long\def\abs#1{\left|#1\right|}

\global\long\def\grad{\M{\nabla}}
\global\long\def\avv#1{\langle#1\rangle}
\global\long\def\av#1{\left\langle #1\right\rangle }

\global\long\def\myhalf{\sfrac{1}{2}}
\global\long\def\mythreehalves{\sfrac{3}{2}}

We performed computer simulations of the experimental setup using finite-volume
methods for fluctuating hydrodynamics described in more detail elsewhere
\cite{LLNS_Staggered,DFDB,MultiscaleIntegrators}; here we summarize
some key points. In particular, Section V.A of the work of Delong
et al. \cite{MultiscaleIntegrators} present simulations of giant
fluctuations in the GRADFLEX experiment, which are used as a basis
for the more detailed computations reported in this work. The numerical
methods have been implemented in the IBAMR software framework \cite{IBAMR}.
We will use CGS units in what follows.
In the numerical computations we align the gradient with the y axes 
in order to unify the notation for two and three dimensional simulations.

Our numerical codes solve the following stochastic partial differential
equations for the fluctuating fluid velocity field $\V v\left(\V r,t\right)$,
the mass concentration $c\left(\V r,t\right)$, and the temperature
$T\left(\V r,t\right)$ \cite{booksengers06}, 
\begin{align}
\rho\partial_{t}\V v+\grad\pi= & \eta\grad^{2}\V v+\grad\cdot\left(\sqrt{2\eta k_{B}T_{0}}\,\M{\mathcal{W}}\right)\label{eq:LLNS_comp_v_simp}\\
\grad\cdot\V v= & 0\nonumber \\
\partial_{t}c+\V v\cdot\grad c= & D\grad\cdot\left(\grad c+c\left(1-c\right)S_{T}\grad T\right)\label{eq:LLNS_comp_c_simp}\\
\partial_{t}T+\V v\cdot\grad T= & \kappa\grad^{2}T,\label{eq:tempr_eq}
\end{align}
where $\M{\mathcal{W}}\left(\V r,t\right)$ denotes white-noise stochastic
forcing driving the thermal fluctuations in the momentum flux (stochastic
stress). Here $\eta=5.21\cdot10^{-3}$ is the shear viscosity, $\pi\left(\V r,t\right)$
is the mechanical pressure, $T_{0}=298$ is the average temperature,
$S_{T}=6.49\cdot10^{-2}$ is the Soret coefficient, $D=1.97\cdot10^{-6}$ is
the diffusion coefficient, and $\kappa=8.95\cdot10^{-4}$ is the
thermal diffusivity. The boundary conditions
for the velocity are no-slip on the bottom and top sapphire walls,
while the other directions are periodic. We will discuss boundary
conditions for temperature and concentration shortly.

A number of physical approximations have been made in formulating
the system of equations (\ref{eq:LLNS_comp_v_simp},\ref{eq:LLNS_comp_c_simp},\ref{eq:tempr_eq}).
First, we have ignored thermal fluctuations in the mass flux and in
the heat flux, which are responsible for equilibrium fluctuations
in the concentration and temperature; this is justified since our
focus is on the much larger non-equilibrium fluctuations. Second, we
have used a constant temperature $T_{0}$ for the stochastic stress
tensor instead of a spatially-varying temperature; this is justified
because the maximum difference in temperature across the sample is
on the order of a tens of degrees. Third, the density $\rho=0.858$
is taken to be constant in a Boussinesq approximation. 

In linearized fluctuating hydrodynamics the equations (\ref{eq:LLNS_comp_v_simp},\ref{eq:LLNS_comp_c_simp},\ref{eq:tempr_eq})
are expanded to leading order in the magnitude of the fluctuations
$\d c=c-\av c$, $\d T=T-\av T$ and $\d{\V v}=\V v-\av{\V v}=\V v$
around the steady state solution of the deterministic equations \cite{booksengers06}.
As explained in detail in Ref. \citenum{MultiscaleIntegrators}, our
numerical methods perform this linearization numerically by solving

the fully nonlinear equations with weak noise. For the example studied
here, in the linearized fluctuating hydrodynamics regime, there is
no difference between two and three-dimensional simulations due to
the symmetries of the problem. Also note that in microgravity the
temperature and concentration fluctuations are completely decoupled
since there is no buoyancy terms feeding back into the momentum (velocity)
equation. Therefore, numerically we separately solve (\ref{eq:LLNS_comp_v_simp},\ref{eq:LLNS_comp_c_simp})
for concentration when examining concentration fluctuations, and we
separately solve (\ref{eq:LLNS_comp_v_simp},\ref{eq:tempr_eq}) when
examining temperature fluctuations. The reason for this is that these
two cases require different temporal integrators, as explained in
extensive detail in Ref. \citenum{MultiscaleIntegrators}. We therefore
separately discuss concentration and temperature fluctuations.

The experimentally observed light intensity, once corrected for the
optical transfer function of the equipment, is proportional to the
intensity of the fluctuations in the concentration and temperature
averaged along the gradient. The contribution due to concentration
fluctuations to the shadowgraph is therefore related to the Fourier
transform $\widehat{c}_{\perp}\left(k,t\right)$ of the vertically
averaged concentration, 
$
c_{\perp}(x,z;t)=L^{-1}\int_{0}^{L}c(x,y,z;t)dy,
$
where $L=0.1$ is the thickness of the sample. More specifically,
our simulations compute the time-dependent static structure factor
$
S\left(k_{x},k_{z};t\right)=\av{\left(\widehat{\d c}_{\perp}\right)\left(\widehat{\d c}_{\perp}\right)^{\star}},
$
and similarly for temperature fluctuations.

\subsubsection*{Concentration fluctuations}

Typical liquid mixtures have a large Schmidt number, $S_{c}=\nu/D\gg1$,
in particular, for the GRADFLEX mixture $S_{c}\approx3\cdot10^{3}$.
This makes direct numerical solution of the original \emph{inertial}
equations (\ref{eq:LLNS_comp_v_simp},\ref{eq:LLNS_comp_c_simp})
numerically infeasible; the time step size needs to be chosen to resolve
vorticity fluctuations but the time scale of interest is the much
longer mass diffusion time scale. Therefore, we first take a limit
of equations (\ref{eq:LLNS_comp_v_simp},\ref{eq:LLNS_comp_c_simp})
as $S_{c}\rightarrow\infty$; in the linearized setting this \emph{overdamped}
limit amounts to deleting the inertial term $\rho\partial_{t}\V v$
in the velocity equation \cite{MultiscaleIntegrators}. Lastly, it
is convenient to approximate the Soret flux $c\left(1-c\right)S_{T}$
with the linearization $cS_{T}$, which is valid since $c\ll1$; this
helps us treat this term implicitly in our numerical methods and thus
strictly conserve mass. 

In summary, concentration fluctuations are modeled using the equations,
in addition to incompressibility,
\begin{align}
\grad\pi= & \eta\grad^{2}\V v+\grad\cdot\left(\sqrt{2\eta k_{B}T_{0}}\,\M{\mathcal{W}}\right)\label{eq:LLNS_comp_v_simp-1}\\
\partial_{t}c+\V v\cdot\grad c= & D\grad\cdot\left(\grad c+cS_{T}\grad T\right).\label{eq:LLNS_comp_c_simp-1}
\end{align}
The boundary conditions on the top and bottom boundaries (sapphire
plates) are zero flux boundary conditions, giving the Robin boundary
condition $\grad c=-cS_{T}\grad T$ at the boundaries. The initial
condition we start from is a uniform solution of concentration $c_{0}=0.018$;
with time this decays to an exponential average profile that solves
$\frac{d}{dy}\av c=-\av cS_{T}\grad T$  (see Fig. \ref{fig concprof}). In our simulations we have accounted
for the initial transient in establishing the concentration profile
across the sample. Based on measurements of the time
response of the PID servos that control the temperature of the sapphire windows, the temperature gradient in the $y$ direction is modeled
with the following empirical fit as a function of time,
$ \grad T=\frac{\D T}{L} \left[1-\exp\left(0.00540\, t-0.0602\, t^{0.816}\right)\right]\hat{\V y}$,
where $\D T=13.25$ is the estimated steady-state temperature difference.

The spatial discretization are essentially identical to those in our
previous work \cite{LLNS_Staggered}. The simulations of the transient
development of concentration fluctuations used the overdamped temporal
integrator summarized in Algorithm 3 in Ref. \citenum{MultiscaleIntegrators}. We perform fully three-dimensional simulations on a domain of dimensions
$0.654\times0.1\times0.654$ (this tries to match the smallest wavenumber
in the simulations with the wave numbers measured with CCD camera in
the experiments), discretized on a $256\times40\times256$ grid, using
a time step size of $\D t=10\,\mbox{s}$. The structure factors $S\left(k_{x},k_{z};t\right)$
were averaged radially to obtain $S\left(k;t\right)$, where $k=\sqrt{k_{x}^{2}+k_{z}^{2}}$,
using an averaging procedure that mimics that used in the analysis of the experimental data.
Note that in this case it is possible to obtain the same results using
two-dimensional simulations ($k_{z}=0$) because of the symmetries
of the linearized equations. Nevertheless, we chose to obtain three-dimensional
results directly comparable to experiments. Sixteen independent simulations
were performed and the results averaged to reduce statistical noise
and estimate statistical error bars. To obtain the static structure
factor at steady state, we initialized the system using the steady
state concentration profile, and fixed $\nabla T=\D T/L$. For these
steady-state runs we used a time step size $\D t=80$s and averaged
over a single run of 2000 time steps (corresponding to about 44 hours of physical time)
skipping the initial 200 time steps in the analysis in order to allow
the system time to reach a statistical steady state.

\subsubsection*{Temperature fluctuations}

The dynamics of velocity and temperature (\ref{eq:LLNS_comp_v_simp},\ref{eq:tempr_eq})
occur at similar time scales and must be integrated together; it is
not justified to delete the inertial term $\rho\partial_{t}\V v$
in the velocity equation as it was for concentration. Therefore, for
temperature we solve the system of equations
\begin{align}
\rho\partial_{t}\V v+\grad\pi= & \eta\grad^{2}\V v+\grad\cdot\left(\sqrt{2\eta k_{B}T_{0}}\,\M{\mathcal{W}}\right)\label{eq:LLNS_comp_v_simp-2}\\
\partial_{t}T+\V v\cdot\grad T= & \kappa\grad^{2}T.\label{eq:tempr_eq-1}
\end{align}
The boundary condition for temperature at the top and bottom walls
are Dirichlet conditions, with $T(y=0,t)=304.6$ at one of the boundaries,
and $T(y=L,t)=291.4$ at the other wall, leading to a linear steady
state temperature profile. Since for temperature we are not interested
in the transient behavior, but rather only the steady state static
structure factor, the initial temperature field is set to be the linear
steady state.

The spatial discretization is identical to that for concentration,
in fact, our computer code does not distinguish between temperature
and concentration since the equations are essentially identical. The
temporal integrator is the inertial scheme summarized in Algorithm
1 in Ref. \citenum{MultiscaleIntegrators}, requiring a much smaller
time step size $\D t=0.0016$ s in order to resolve the fast vorticity dynamics.
In this case we perform two dimensional simulations in a domain of
dimensions $0.8\times0.1$ on a grid of $256\times32$ grid cells.
We average over 16 simulations of $5\cdot10^{5}$ time steps each
(corresponding to about 800 s of physical time), skipping
the initial $5\cdot10^{4}$ time steps.

\section*{Acknowledgements}

We thank D. S. Cannell, M. Giglio, S. Mazzoni, C. J. Takacs, O. Minster, A. Verga, F. Molster, N. Melville, W. Meyer, A. Smart, R. Greger, B. Hirtz, and R. Pereira for their contribution to the GRADFLEX project. We are indebted to F. Giavazzi for help with the analysis of results, to J. M. Ortiz de Zarate for providing us the results of his exact theoretical model, and to B. Griffith for developing the IBAMR software used to perform the simulations reported here. We acknowledge the contribution of the Telesupport team and of the industrial consortium led by RUAG aerospace. Ground-based activity was supported by ESA and NASA. Flight opportunity sponsored by ESA. A. D. was funded in part by the U.S. DOE ASCR program under Award Number DE-SC0008271, and by the U.S. NSF under grant DMS-1115341.


\begin{thebibliography}{10}
\expandafter\ifx\csname url\endcsname\relax
  \def\url#1{\texttt{#1}}\fi
\expandafter\ifx\csname urlprefix\endcsname\relax\def\urlprefix{URL }\fi
\providecommand{\bibinfo}[2]{#2}
\providecommand{\eprint}[2][]{\url{#2}}

\bibitem{booksengers06}
\bibinfo{author}{Ortiz~de Z\'arate, J.~M.} \& \bibinfo{author}{Sengers, J.~V.}
\newblock \emph{\bibinfo{title}{Hydrodynamic Fluctuations in Fluids and Fluid
  Mixtures}} (\bibinfo{publisher}{Elsevier}, \bibinfo{year}{2006}).

\bibitem{grinstein95}
\bibinfo{author}{Grinstein, G.}
\newblock \bibinfo{title}{Generic scale invariance and self-organized
  criticality}
\newblock in 
  \emph{\bibinfo{booktitle}{Scale Invariance, Interfaces, and Non-Equilibrium
  Dynamics}}, 
  \bibinfo{editor}{(ed McKane, A. \emph{et~al.})}
  \bibinfo{pages}{261--293} (\bibinfo{publisher}{Plenum},
  \bibinfo{year}{2006}).

\bibitem{brogioli00}
\bibinfo{author}{Brogioli, D.} \& \bibinfo{author}{Vailati, A.}
\newblock \bibinfo{title}{Diffusive mass transfer by nonequilibrium
  fluctuations: Fick's law revisited}.
\newblock \emph{\bibinfo{journal}{Phys. Rev. E}} \textbf{\bibinfo{volume}{63}},
  \bibinfo{pages}{02105--1--4} (\bibinfo{year}{2000}).

\bibitem{donev11}
\bibinfo{author}{Donev, A.}, \bibinfo{author}{de~la Fuente, A.},
  \bibinfo{author}{Bell, J.~B.} \& \bibinfo{author}{Garcia, A.~L.}
\newblock \bibinfo{title}{Diffusive transport enhanced by thermal velocity
  fluctuations}.
\newblock \emph{\bibinfo{journal}{Phys. Rev. Lett.}}
  \textbf{\bibinfo{volume}{106}}, \bibinfo{pages}{204501--1--4}
  (\bibinfo{year}{2011}).

\bibitem{donev14}
\bibinfo{author}{Donev, A.}, \bibinfo{author}{Fai, T.~G.} \&
  \bibinfo{author}{Vanden-Eijnden, E.}
\newblock \bibinfo{title}{A reversible mesoscopic model of diffusion in
  liquids: from giant fluctuations to fick's law}.
\newblock \emph{\bibinfo{journal}{J. Stat. Mech}}
  \textbf{\bibinfo{volume}{P04004}}, \bibinfo{pages}{1--39}
  (\bibinfo{year}{2014}).

\bibitem{segre93}
\bibinfo{author}{Segr\'e, P.~N.} \& \bibinfo{author}{Sengers, J.~V.}
\newblock \bibinfo{title}{Nonequilibrium fluctuations in liquid mixtures under
  the influence of gravity}.
\newblock \emph{\bibinfo{journal}{Physica A}} \textbf{\bibinfo{volume}{198}},
  \bibinfo{pages}{46--77} (\bibinfo{year}{1993}).

\bibitem{vailati96}
\bibinfo{author}{Vailati, A.} \& \bibinfo{author}{Giglio, M.}
\newblock \bibinfo{title}{q divergence of nonequilibrium fluctuations and its
  gravity-induced frustration in a temperature stressed liquid mixture}.
\newblock \emph{\bibinfo{journal}{Phys. Rev. Lett.}}
  \textbf{\bibinfo{volume}{77}}, \bibinfo{pages}{1484--1487}
  (\bibinfo{year}{1996}).

\bibitem{vailati97}
\bibinfo{author}{Vailati, A.} \& \bibinfo{author}{Giglio, M.}
\newblock \bibinfo{title}{Giant fluctuations in a free diffusion process}.
\newblock \emph{\bibinfo{journal}{Nature}} \textbf{\bibinfo{volume}{390}},
  \bibinfo{pages}{262--265} (\bibinfo{year}{1997}).

\bibitem{wu95}
\bibinfo{author}{Wu, M.}, \bibinfo{author}{Ahlers, G.} \&
  \bibinfo{author}{Cannell, D.~S.}
\newblock \bibinfo{title}{Thermally induced fluctuations below the onset of
  rayleigh-bénard convection}.
\newblock \emph{\bibinfo{journal}{Phys. Rev. Lett.}}
  \textbf{\bibinfo{volume}{75}}, \bibinfo{pages}{1743--1746}
  (\bibinfo{year}{1995}).

\bibitem{ahlers}
\bibinfo{author}{Oh, J.}, \bibinfo{author}{Ortiz~de Z\'arate, J.~M.},
  \bibinfo{author}{Sengers, J.~V.} \& \bibinfo{author}{Ahlers, G.}
\newblock \bibinfo{title}{Dynamics of fluctuations in a fluid below the onset
  of Rayleigh-Bénard convection}.
\newblock \emph{\bibinfo{journal}{Phys. Rev. E}} \textbf{\bibinfo{volume}{69}},
  \bibinfo{pages}{021106--1--13} (\bibinfo{year}{2004}).

\bibitem{giavazzi09}
\bibinfo{author}{Giavazzi, F.} \& \bibinfo{author}{Vailati, A.}
\newblock \bibinfo{title}{Scaling of the spatial power spectrum of excitations
  at the onset of solutal convection in a nanofluid far from equilibrium}.
\newblock \emph{\bibinfo{journal}{Phys. Rev. E}} \textbf{\bibinfo{volume}{80}},
  \bibinfo{pages}{015303--1--4(R)} (\bibinfo{year}{2009}).

\bibitem{vailati2011}
\bibinfo{author}{Vailati, A.} \emph{et~al.}
\newblock \bibinfo{title}{Fractal front of diffusion in microgravity}.
\newblock \emph{\bibinfo{journal}{Nat. Commun.}} \textbf{\bibinfo{volume}{2}},
  \bibinfo{pages}{290} (\bibinfo{year}{2011}).

\bibitem{vailati98}
\bibinfo{author}{Vailati, A.} \& \bibinfo{author}{Giglio, M.}
\newblock \bibinfo{title}{Nonequilibrium fluctuations in time dependent
  diffusion processes}.
\newblock \emph{\bibinfo{journal}{Phys. Rev. E}} \textbf{\bibinfo{volume}{58}},
  \bibinfo{pages}{4361--4371} (\bibinfo{year}{1998}).

\bibitem{proteins}
\bibinfo{author}{De~Lucas, L.~J. \emph{et~al.}}
\newblock \bibinfo{title}{Protein crystal growth in microgravity}.
\newblock \emph{\bibinfo{journal}{Science}} \textbf{\bibinfo{volume}{246}},
  \bibinfo{pages}{651--654} (\bibinfo{year}{1989}).

\bibitem{snell05}
\bibinfo{author}{Snell, E.~H.} \& \bibinfo{author}{Helliwell, J.~R.}
\newblock \bibinfo{title}{Macromolecular crystallization in microgravity}.
\newblock \emph{\bibinfo{journal}{Rep. Prog. Phys.}}
  \textbf{\bibinfo{volume}{68}}, \bibinfo{pages}{799--853}
  (\bibinfo{year}{2005}).

\bibitem{barmatz07}
\bibinfo{author}{Barmatz, M.}, \bibinfo{author}{Hahn, I.},
  \bibinfo{author}{Lipa, J.~A.} \& \bibinfo{author}{Duncan, R.~V.}
\newblock \bibinfo{title}{Critical phenomena in microgravity: past, present and
  future}.
\newblock \emph{\bibinfo{journal}{Rev. Mod. Phys}}
  \textbf{\bibinfo{volume}{79}}, \bibinfo{pages}{1--52} (\bibinfo{year}{2007}).

\bibitem{beysens14}
\bibinfo{author}{Beysens, D.}
\newblock \bibinfo{title}{Critical point in space: a quest for universality}.
\newblock \emph{\bibinfo{journal}{Microgravity Sci. Tec.}}
  \textbf{\bibinfo{volume}{26}}, \bibinfo{pages}{201--218}
  (\bibinfo{year}{2014}).

\bibitem{shevtsova10}
\bibinfo{author}{Shevtsova, V.}
\newblock \bibinfo{title}{Ividil experiment onboard the iss}.
\newblock \emph{\bibinfo{journal}{Adv. Space Res.}}
  \textbf{\bibinfo{volume}{46-51}}, \bibinfo{pages}{672}
  (\bibinfo{year}{2010}).

\bibitem{shevtsova11}
\bibinfo{author}{Shevtsova, V.} \emph{et~al.}
\newblock \bibinfo{title}{Ividil experiment onboard iss: thermodiffusion in
  presence of controlled vibrations}.
\newblock \emph{\bibinfo{journal}{C. R. M\'ecanique}}
  \textbf{\bibinfo{volume}{339}}, \bibinfo{pages}{310--317}
  (\bibinfo{year}{2011}).

\bibitem{shevtsova14}
\bibinfo{author}{Shevtsova, V.}  \emph{et~al.}
\newblock \bibinfo{title}{Diffusion and soret in ternary mixtures. preparation
  of the dcmix2 experiment on the iss}.
\newblock \emph{\bibinfo{journal}{Microgravity Sci. Tec.}}
  \textbf{\bibinfo{volume}{25}}, \bibinfo{pages}{275--283}
  (\bibinfo{year}{2014}).

\bibitem{bookdegroot62}
\bibinfo{author}{de~Groot, S.~R.} \& \bibinfo{author}{Mazur, P.}
\newblock \emph{\bibinfo{title}{Nonequilibrium Thermodynamics}}
  (\bibinfo{publisher}{North-Holland}, \bibinfo{year}{1962}).

\bibitem{booksettles01}
\bibinfo{author}{Settles, G.~S.}
\newblock \emph{\bibinfo{title}{Schlieren and Shadowgraph Techniques}}
  (\bibinfo{publisher}{Springer}, \bibinfo{year}{2001}).

\bibitem{trainoff02}
\bibinfo{author}{Trainoff, S.} \& \bibinfo{author}{Cannell, D.~S.}
\newblock \bibinfo{title}{Physical optics treatment of the shadowgraph}.
\newblock \emph{\bibinfo{journal}{Phys. Fluids}} \textbf{\bibinfo{volume}{14}},
  \bibinfo{pages}{1340--1363} (\bibinfo{year}{2002}).

\bibitem{MultiscaleIntegrators}
\bibinfo{author}{Delong, S.}, \bibinfo{author}{Sun, Y.}, \bibinfo{author}{Griffith, B.~E.}, \bibinfo{author}{Vanden-Eijnden, E.} \&   \bibinfo{author}{Donev, A.}
\newblock \bibinfo{title}{Multiscale temporal integrators for fluctuationg
  hydrodynamics}.
\newblock \emph{\bibinfo{journal}{Phys. Rev. E}} \textbf{\bibinfo{volume}{90}},
  \bibinfo{pages}{063312--1--23} (\bibinfo{year}{2014}).

\bibitem{LLNS_Staggered}
\bibinfo{author}{Balboa~Usabiaga, F.} \emph{et~al.}
\newblock \bibinfo{title}{Staggered schemes for fluctuating hydrodynamics}.
\newblock \emph{\bibinfo{journal}{SIAM J. Multiscale Model. Simul.}}
  \textbf{\bibinfo{volume}{10}}, \bibinfo{pages}{1369--1408} (\bibinfo{year}{2012}).

\bibitem{dezarate04}
\bibinfo{author}{Ortiz~de Z\'arate, J.~M.}, \bibinfo{author}{Peluso, F.} \&
  \bibinfo{author}{Sengers, J.~V.}
\newblock \bibinfo{title}{Nonequilibrium fluctuations in the Rayleigh-Bénard
  problem for binary fluid mixtures}.
\newblock \emph{\bibinfo{journal}{Eur. Phys. J. E}}
  \textbf{\bibinfo{volume}{15}}, \bibinfo{pages}{319--333}
  (\bibinfo{year}{2004}).

\bibitem{spinodalhuang}
\bibinfo{author}{Huang, J.~S.}, \bibinfo{author}{Goldburg, W.~I.} \&
  \bibinfo{author}{Bjierkaas, A.~W.}
\newblock \bibinfo{title}{Study of phase separation in a critical binary liquid
  mixture: spinodal decomposition}.
\newblock \emph{\bibinfo{journal}{Phys. Rev. Lett.}}
  \textbf{\bibinfo{volume}{32}}, \bibinfo{pages}{921--923}
  (\bibinfo{year}{1974}).

\bibitem{spinodalbinder}
\bibinfo{author}{Binder, K.} \& \bibinfo{author}{Stauffer, D.}
\newblock \bibinfo{title}{Theory for the slowing down of the relaxation and
  spinodal decomposition of binary mixtures}.
\newblock \emph{\bibinfo{journal}{Phys. Rev. Lett.}}
  \textbf{\bibinfo{volume}{33}}, \bibinfo{pages}{1006--1009}
  (\bibinfo{year}{1974}).

\bibitem{spinodalmarro}
\bibinfo{author}{Marro, J.}, \bibinfo{author}{Lebowitz, J.~L.} \&
  \bibinfo{author}{Kalos, M.~H.}
\newblock \bibinfo{title}{Computer simulation of the time evolution of a
  quenched model alloy in the nucleation regime}.
\newblock \emph{\bibinfo{journal}{Phys. Rev. Lett.}}
  \textbf{\bibinfo{volume}{43}}, \bibinfo{pages}{282--285}
  (\bibinfo{year}{1979}).

\bibitem{spinodalfurukawa}
\bibinfo{author}{Furukawa, H.}
\newblock \bibinfo{title}{A dynamic scaling assumption for phase separation}.
\newblock \emph{\bibinfo{journal}{Adv. Phys.}} \textbf{\bibinfo{volume}{34}},
  \bibinfo{pages}{703--750} (\bibinfo{year}{1985}).

\bibitem{takacs2011}
\bibinfo{author}{Takacs, C.~J.} \emph{et~al.}
\newblock \bibinfo{title}{Thermal fluctuations in a layer of $CS_2$ subjected
  to temperature gradients with and without the influence of gravity}.
\newblock \emph{\bibinfo{journal}{Phys. Rev. Lett.}}
  \textbf{\bibinfo{volume}{106}}, \bibinfo{pages}{244502--1--4}
  (\bibinfo{year}{2011}).

\bibitem{ruckenstein81}
\bibinfo{author}{Ruckenstein, E.}
\newblock \bibinfo{title}{Can phoretic motions be treated as interfacial
  tension gradient driven phenomena}.
\newblock \emph{\bibinfo{journal}{J. Colloid Interface Sci.}}
  \textbf{\bibinfo{volume}{83}}, \bibinfo{pages}{77--81}
  (\bibinfo{year}{1981}).

\bibitem{croccolo07}
\bibinfo{author}{Croccolo, F.}, \bibinfo{author}{Brogioli, D.}, \bibinfo{author}{Vailati, A.}, \bibinfo{author}{Giglio, M.}, \& \bibinfo{author}{CAnnell, D.~S.} 
\newblock \bibinfo{title}{Non-diffusive decay of gradient driven fluctuations
  in a free-diffusion process}.
\newblock \emph{\bibinfo{journal}{Phys. Rev. E}} \textbf{\bibinfo{volume}{76}},
  \bibinfo{pages}{041112--1--9} (\bibinfo{year}{2007}).

\bibitem{dezarate06}
\bibinfo{author}{Ortiz~de Z\'arate, J.~M.}, \bibinfo{author}{Fornes, J.~A.} \&
  \bibinfo{author}{Sengers, J.~V.}
\newblock \bibinfo{title}{Long-wavelength nonequilibrium concentration
  fluctuations induced by the soret effect}.
\newblock \emph{\bibinfo{journal}{Phys. Rev. E}} \textbf{\bibinfo{volume}{74}},
  \bibinfo{pages}{046305--1--11} (\bibinfo{year}{2006}).

\bibitem{ortizpersonal}
 \bibinfo{author}{Ortiz~de Z\'arate}, \bibinfo{author}{Kirkpatrick, T.~R.} \& \bibinfo{author}{Sengers, J.~V.} 
\newblock \bibinfo{title}{Non-equilibrium concentration fluctuations in binary liquids with realistic boundary conditions}.
\bibinfo{note} {arXiv:1505.01355v1} (\bibinfo{year}{2015}) 


\bibitem{spinodalcarpineti}
\bibinfo{author}{Carpineti, M.} \& \bibinfo{author}{Giglio, M.}
\newblock \bibinfo{title}{Spinodal-type dynamics in fractal aggregation of
  colloidal clusters}.
\newblock \emph{\bibinfo{journal}{Phys. Rev. Lett.}}
  \textbf{\bibinfo{volume}{68}}, \bibinfo{pages}{3327--3330}
  (\bibinfo{year}{1992}).

\bibitem{vailati06}
\bibinfo{author}{Vailati, A.} \emph{et~al.}
\newblock \bibinfo{title}{Gradient-driven fluctuations experiment: fluid
  fluctuations in microgravity}.
\newblock \emph{\bibinfo{journal}{Applied Optics}}
  \textbf{\bibinfo{volume}{45}}, \bibinfo{pages}{2155--2165}
  (\bibinfo{year}{2006}).

\bibitem{cerbino08}
\bibinfo{author}{Cerbino, R.} \emph{et~al.}
\newblock \bibinfo{title}{X-ray-scattering information obtained from near-field
  speckle}.
\newblock \emph{\bibinfo{journal}{Nature Phys.}} \textbf{\bibinfo{volume}{4}},
  \bibinfo{pages}{238--243} (\bibinfo{year}{2008}).
  
\bibitem{cocis}
\bibinfo{author}{Cerbino, R.} \& \bibinfo{author}{Vailati A.}
\newblock \bibinfo{title}{Near-field scattering techniques: Novel instrumentation and results from time and spatially resolved investigations of soft matter systems}
\newblock \emph{\bibinfo{journal}{Curr. Op. Coll. Int. Science}} \textbf{\bibinfo{volume}{14}},
  \bibinfo{pages}{416--425} (\bibinfo{year}{2009}).
  
  \bibitem{java}
\bibinfo{author}{Giavazzi, F.} \& \bibinfo{author}{Cerbino, R.}
\newblock \bibinfo{title}{Digital Fourier Microscopy for Soft Matter Dynamics}
\newblock \emph{\bibinfo{journal}{J. Opt.}} \textbf{\bibinfo{volume}{16}},
  \bibinfo{pages}{083001} (\bibinfo{year}{2014}).
  

\bibitem{DFDB}
\bibinfo{author}{Delong, S.}, \bibinfo{author}{Griffith, B.~E.}, \bibinfo{author}{Vanden-Eijnden, E.} \&
  \bibinfo{author}{Donev, A.}
  \newblock \bibinfo{title}{Temporal Integrators for Fluctuating Hydrodynamics}
  \newblock \emph{\bibinfo{journal}{Phys. Rev. E}} \textbf{\bibinfo{volume}{87}},
    \bibinfo{pages}{033302--1--22} (\bibinfo{year}{2013}).

\bibitem{IBAMR}
\bibinfo{author}{Griffith, B.~E.}, \bibinfo{author}{Hornung, R.~D}, \bibinfo{author}{McQueen, D.~M} \&
  \bibinfo{author}{Peskinv, C.~S}
    \newblock \bibinfo{title}{An adaptive, formally second order accurate version of the immersed
      boundary method}
    \newblock \emph{\bibinfo{journal}{J. Comput. Phys.}} \textbf{\bibinfo{volume}{223}},
      \bibinfo{pages}{10--49} (\bibinfo{year}{2007}).
  
  
\end{thebibliography}
\end{document}